# Toward Open Earth Science as Fast and Accessible as Natural Language

**Marquita Ellis**[1], **Iksha Gurung**[2], **Muthukumaran Ramasubramanian**[2], **Rahul Ramachandran**[3]
[1]IBM Research, Almaden, CA, USA.
[2]Earth System Science Center, The University of Alabama in Huntsville, AL, USA.
[3]NASA Marshall Space Flight Center, Huntsville, AL, USA.
**Correspondence:** m.ellis@ibm.com

## Abstract

*Is natural-language-driven earth observation data analysis now feasible with the assistance of Large Language Models (LLMs)?* For open science in service of public interest, *feasibility* requires reliably high accuracy, interactive latencies, low (sustainable) costs, open LLMs, and openly maintainable software — hence, the challenge. *What are the techniques and programming system requirements necessary for satisfying these constraints, and what is the corresponding development and maintenance burden in practice?* This study lays the groundwork for exploring these questions, introducing an impactful earth science use-case, and providing a software framework with evaluation data and metrics, along with initial results from employing model scaling, prompt-optimization, and inference-time scaling optimization techniques. While we attain high accuracy (near 100%) across 10 of 11 metrics, the analysis further considers cost (token-spend), latency, and maintainability across this space of techniques. Finally, we enumerate opportunities for further research, general programming and evaluation framework development, and ongoing work for a comprehensive, deployable solution. This is a call for collaboration and contribution.

## 1 Introduction

The ever-increasing volume, velocity, and veracity of Earth observation data present significant challenges to efficient data discovery and analysis. While systems like the Common Metadata Repository (CMR), SpatioTemporal Asset Catalogs (Newman and ESCO, 2023), and Microsoft Planetary Computer (Microsoft Open Source et al., 2022) facilitate access, they often require specialized knowledge of query languages and data structures, creating a barrier for many researchers, particularly those outside of informatics. This limits the full potential of these valuable datasets for applications in atmospheric science, climate monitoring, policy planning, and emergency response. Large Language Models (LLMs), with their advanced Natural Language Processing (NLP) capabilities, offer a compelling solution to democratize access to this complex data, enabling intuitive query formulation using natural language.

This study investigates the feasibility of a LLM-powered interface for simplified Earth science data retrieval. A core challenge in such a system lies in the accurate interpretation of spatiotemporal relationships expressed in natural language. The system must robustly identify the geographic location, temporal window, and specific data type (e.g., event, observation) requested by the user. This requires not only extracting explicitly stated parameters (area, date, event type) but also inferring implicit constraints and performing sophisticated temporal reasoning, such as interpreting relative time references ("last week," "since 2020"). Furthermore, while the primary focus is on these core parameters, the system implicitly considers, and could be extended to explicitly handle, characteristics like data resolution and sensor type.

Our primary contribution is a novel system that reformulates geospatial data querying as a Named Entity Recognition (NER) task, extracting key parameters (area, date, event type) from natural language queries. We introduce a new, rigorously validated evaluation dataset comprising over 100 query-answer pairs, developed in collaboration with domain scientists. This dataset, validated both manually and through automated checks, provides a benchmark for evaluating the performance of LLM-based geospatial query systems.

Furthermore, we explore and evaluate a range of optimization techniques to enhance system performance. These include model scaling, programmatic prompt optimization using the DSPy framework (Khattab et al., 2024), and inference-time strategies such as self-refinement and task decomposition (Madaan et al., 2024). We analyze the impact of these techniques on accuracy, cost, and latency, with a particular focus on improving temporal reasoning capabilities. The evaluation metrics employed to quantify system performance are detailed in Section 2.5, providing a framework for analyzing the strengths and weaknesses of different approaches.

| Parameter | Definition |
|---|---|
| `area` | The physical location of interest, anywhere on Earth. |
| `date` | The date (range) of interest. |
| `event_type` | A descriptor corresponding to supported analysis types (e.g. flood). |

Table 1: Key parameters for approach validation.

This research demonstrates the significant potential of LLMs to revolutionize access to Earth science data, empowering a wider range of users to leverage this information for critical scientific and societal applications.

## 2 Approach

### 2.1 Problem Formulation

In order to retrieve and analyze earth observation data, past approaches require users to learn and employ specialized query languages and data structures. This constitutes a barrier for many researchers, particularly those outside informatics (Section 1). Our approach reformulates geospatial data querying as a Named Entity Recognition (NER) task. Instead of code, the user states queries in natural language, from which the system captures parameters for driving analysis and delivering processed images over the specified time scale.

This study explores and validates this approach in light of its simplicity and extensibility, including the ease of adding more powerful features, e.g. speech vs text driven analysis. Table 2.1 presents the minimal set of parameters identified for concept validation. `Area` and `date`(s) are required for image retrieval. The `event_type` corresponds to the analysis types supported via integrated Prithvi geospatial foundation models (Jakubik et al., 2023). Additional properties, such as bounding boxes, are derivable from these parameters. The parameters can also be easily extended to include e.g. sensor type, image granularity, etc.

### 2.2 Evaluation Data Set Creation

We developed an original evaluation data set for three main reasons: (1) systematic optimization of the overall system design (prompts, model choices, composition, etc.); (2) improving the system over time with grounded evaluation; (3) gaining confidence that the system meets end-user needs in practice, with data representing actual user (domain scientist) queries and accepted answers.

The evaluation data currently consists of over 100 unique query-answer pairs that were manually crafted with domain scientists and that underwent both manual and automated validation. The answers are JSON translations of representative user queries. Prior to this work, no standardized data set existed for this task. The following is an example.

```
Query: "July 14, 2023, flooding in Seoul"
Answer: {"area": "Seoul", "date": "2023-07-14",
    "event_type":"flood"}
```

The queries vary in complexity, particularly in the dimension of temporal reference interpretation. In certain queries, such as the first example, the temporal window of interest is explicitly given and simply requires `YYYY-MM-DD` reformatting. Other queries contain relative time references, such as "yesterday", "this Friday", or "from the past week", which require not only identifying the pertinent words for date extraction, but also distinguishing the time of the query from that of its subject and reasoning about the distance and span. For simplicity, the first date in a time window of interest is preferred. See the following example.

```
Query: "Find burn scars in the Andes Mountains
    from last season."
Answer: {"area": "Andes Mountains", "date
    ":"2024-03-01", "event_type":"burn_scars"}
```

Limitations of the current dataset version are discussed in Section 6.3.

### 2.3 Semi-Automatic Data Set Refinement

As part of the generation performance evaluation, we incorporated *LLM-as-a-Judge* (Zheng et al., 2023). Given an answer, the LLM-judge is prompted to determine whether the generated answer, particularly whether its date(s), are consistent with the user's query and to provide justification. We submitted the golden answers and queries to the LLM-judge. Rather than catching erroneous or problematic judgments in the query-answer pairs that had been manually vetted multiple times, we found the LLM-judge correctly identified mistakes in the golden answers and ambiguities in the queries. Manually considering the LLM-judge's rationale alongside the answers flagged for inconsistency, we caught and corrected 4 off-by-one date calculations, and 42 QA pairs embedding an assumption of "today's date", decipherable only from the golden answer. Despite the data set being relatively small (just over 100 QA pairs), this additional layer of automatic validation can facilitate systematic validation over time as the data set expands. For deterministic evaluation, queries with relative time references were augmented with a suffix, "Today is…", where today was calculated relative to the accepted answer. The following is one example. In a deployment setting, today's date may similarly be injected into the prompt, but this is one (static versus dynamic) feature distinguishing the evaluation data from real-time user input.

```
Query: "Provide the latest imagery of flooding
     in Houston, Texas, from this past Tuesday.
     Today is June 4, 2024."
Answer: {"area": "Houston, Texas", "date":
     "2024-05-28","event_type":"flood"}
```

### 2.4 Task Complexity

Considering the domain and range of valid input and output values is helpful for understanding the complexity of each subtask. The range of event types corresponds to the system's supported analysis types –currently powered by Privthi models (Jakubik et al., 2023; IBM and NASA, 2023). Hence as an independent practical limitation, the set cardinality is finite and relatively small. English references, variations and synonyms for these event types are also relatively limited. Recognizing the event type of interest in a user query is therefore the most constrained task, essentially multiple choice question answering. By contrast for area recognition, the domain and range is "any physical location on Earth". The evaluation data set currently includes nouns for location references, but a user may submit numerical coordinates.

While filtering invalid queries is delegated to an external router, queries lacking sufficient specificity may be permitted. The language model is prompted not only to recognize the entities named in the query, but also to identify and explain errors encountered in processing the query. This presents a challenge both for generation and verification. For verification, identifying whether an expected error message is present (or not) may be simple, but deciding whether the error message supplied is valid is not as trivial.

Lastly, the temporal reasoning required varies significantly across queries as described in Section 2.2. Moreover, the valid domain and range are limited to the past, and the range is limited to dates or lists of dates that can be represented in the format, `YYYY-MM-DD`, but is otherwise unlimited.

### 2.5 Evaluation Metric Specification

Initial experiments showed generated answers contained an exact answer-match relatively frequently (55%). However, answer-match was uninformative for understanding the remaining performance gap. Furthermore, it did not represent the primary optimization criteria well, given the range and variability of valid answers, and nuances or implicit assumptions of real-world user queries. (See also Section 2.2.) From initial repeated experiments, we identified several expectations for high quality answers. Most became embedded in the handwritten prompt during rapid prototyping, but not all and not their relative priorities (weights). For systematic evaluation and optimization, we encoded these into 10 easily measurable, deterministic metrics and 1 LLM-assisted metric defined as follows.

1. ***Valid JSON*** ; whether an answer substring is syntactically correct JSON and can be parsed into a JSON object. In light of the multiple (changeable) system components supporting JSON structuring (embedded in e.g. the LLM, the provider's serving backend, etc.), this metric can help ensure valid JSON is consistently produced across software versions.

2. ***Contains Expected Error Message*** ; if any field cannot be extracted from the given query, an error message is expected. This metric determines whether that expectation has been met.

3. ***Valid Key Names*** ; whether keys outside the set `{area, date, event_type, error}`, have been generated. This metric was derived from early observations of LLM generated answers including erroneous key names.

4. ***All Required Keys Present*** ; whether all of the requested keys (`area, date, event_type`) are included in the answer. This serves as a minimal indicator that the components of the task are clear even if the task is not completed as desired.

5. ***Valid Event Type*** ; whether the answer's event type is in the set of supported event types (`flood, burn_scars, crops`). Early observations of LLM generated answers included creative alternatives, unsupported by the overall system.

6-7. ***Equivalent Event and Area Values*** ; if key values do not exactly match the golden answer, are they equivalent within a parametrized degree of freedom? Normalized insertion-deletion similarity is used as a measure equivalence. This is applied to both area and event-type values.

8-9. ***Consistent Event and Area Values*** ; whether the area and event type in the answer are mentioned in the user query. This metric was derived from early observations of LLM generated answers including areas and events not mentioned in the original query. Given the constraints on possible answers, especially for event types, implementation using synonyms, string manipulation, and string similarity was straightforward and effective.

10. ***Date Equivalence*** ; whether date(s) generated are numerically equivalent to the golden answer's.

11. ***Date Consistency*** ; whether the answer's date(s) can reasonably be derived from the original user query, independent of (in)equivalence to the golden answer. The implementation employs an LLM to make the determination and provide corresponding rationale.

Using these metrics in combination, we were able to construct meaningful approximations of exact-match and instruction-following, effectively distinguishing semantic and syntactic errors in answer generation. Furthermore, considering results for each metric in isolation revealed the relative difficulties of each subtask, and also measurable trade-offs across models and designs. In our empirical study, the most difficult subtask proved to be date extraction. While tuning the prompt, changing models, model compositions, and otherwise optimizing for date extraction, we were also able to monitor losses in other dimensions via this breakdown of metrics, and thus systematically compare design choices.

### 2.6 Implementation in DSPy

DSPy (Khattab et al., 2024) is an open-source framework for programmatic prompt optimization in Python. Based on its success for other use-cases[1], we decided to see how much the initial handwritten prompt could be improved using DSPy. The results of simply translating the original prompt into a DSPy `Signature` (prior to further prompt optimization) yielded immediate gains. Of perhaps even greater value however, using the framework facilitated implementation of a systematic evaluation process with codified metrics and data in a standardized format.

[1]https://dspy.ai/dspy-usecases/. Last access: Feb 12, 2025.

In the DSPy versions available at the time of this study, we also encountered certain short-comings using DSPy for our use-case, short-comings which influenced the initial implementation of our evaluation system. First, DSPy was initially designed for single-objective optimization. Support for expressing multiple optimization objectives is limited to (1) combining the objectives into a single function that produces a binary "pass" or "fail" result per sample, (2) using DSPy Assertions or Suggestions (Singhvi et al., 2024) to catch-retry-fail when an objective is not met during optimization or runtime or (3) to separately, independently evaluate and optimize for each objective individually. Ultimately, we found (3) to be most informative for ablation studies while not requiring extensive auxiliary code. While (2) works well for its intended purpose, catching and reducing undesirable behavior, it was not designed to yield a performance breakdown across multiple objectives encoded as Suggestion/Assertion invariants. Violation of an invariant either halts the program or must be ignored before subsequent invariants are evaluated. With instrumentation, logging, and configuration specific to our purposes, this limitation could be overcome. However, (3) was the most readily usable approach for understanding the strengths and weaknesses of various design alternatives (prompting techniques, model choices, model compositions, etc.). For multi-objective optimization, all three approaches remain limited. Each objective is implicitly weighted equally.

DSPy was also originally designed for prompt optimization of a single LLM (`module`) at a time in a sequential pipeline. Expressing general program structures with the potential for task-parallel execution is under development[2]. We implemented workarounds to explore inference-time scaling and task decomposition optimizations. However, ongoing work in DSPy and alternative frameworks removing these limitations could reduce manual effort, code volume, and ease of exploiting task-parallelism.

## 3 Optimization Methods and Results

### 3.1 Initial Evaluation and Model Selection

First, we examine the performance of a handwritten prompt developed during system prototyping. To understand attainable performance, we chose a well-vetted general-purpose and open model family with an extreme range of sizes, LLaMA 3.1 (Meta AI, 2024), and varied sizes between the extremes of 405B and 8B. With iterative manual tweaks to the original prompt, we attained 99.065 +/- 0.935% accuracy across models and the 9 metrics unrelated to temporal reference interpretation. The significant performance gaps that remained (up to 35%) were in generating dates equivalent with the golden answers, and/or generating dates consistent with the user query. *Date Equivalence* is determined using deterministic heuristics that supports a single versus multiple valid interpretations of time references in the user query, whereas *Date Consistency* with the user query is determined by a LLM-judge (LLaMA 3.1 405B for all experiments). We needed not only to address this gap but also to formalize the prompt tuning process for software maintainability and extensibility. The following is a small case study highlighting extensibility and robustness concerns with respect to otherwise decoupled system changes.

[2]https://github.com/stanfordnlp/dspy/releases. Last accessed March 10, 2025.

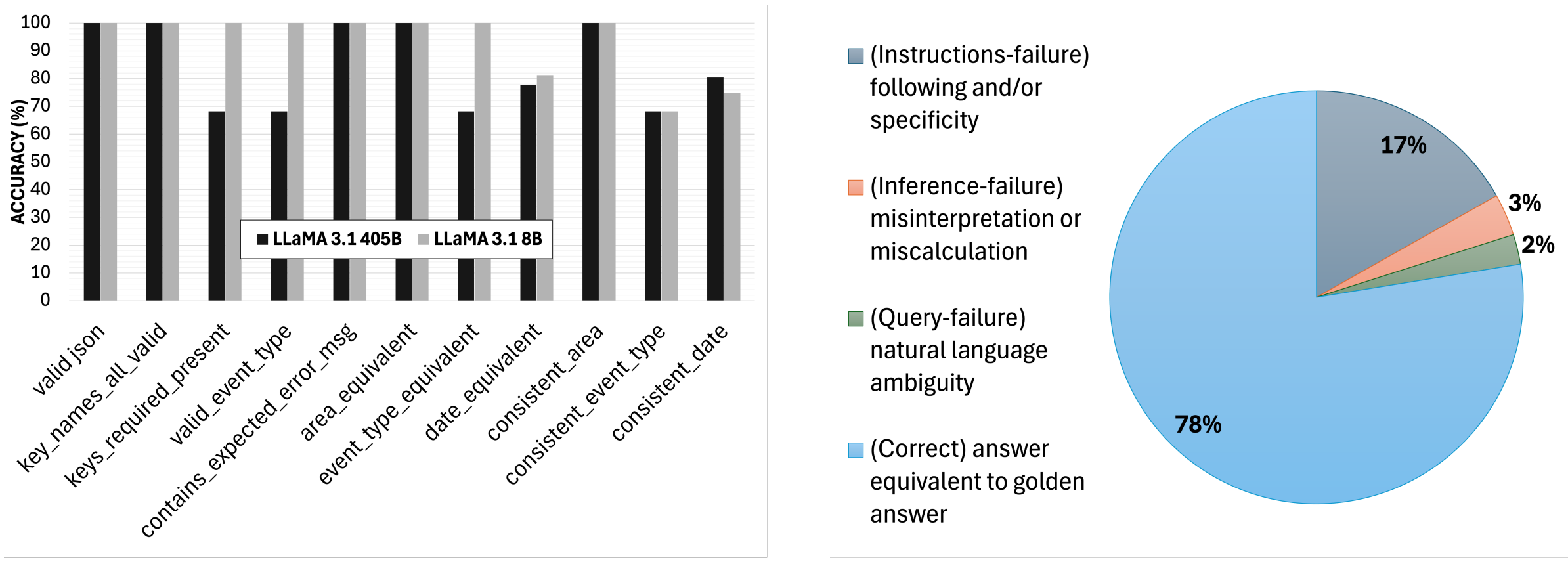

Figure 1: From the case study in Section 3.1, (Left) LLaMA 3.1 405B versus 8B performance across all metrics after introducing a new event type, "crops". (Right) Categorization and percentages of answers from LLaMA 3.1 405B failing *Date Equivalence*.

**Case Study**: Part of the vision for the system is to introduce more analysis types over time, as more geospatial foundation model fine-tunes are integrated. With this in mind, we introduced a third event type, `crops`, into the user query set along with respective golden answers. Figure 1 (Left) shows an interesting difference between LLaMA 3.1 405B and 8B's answer-generation in this scenario. On the surface, it appears LLaMA 3.1 8B is more resilient than 405B to these changes. However, analysis of the traces shows both failed to follow instructions but in different ways. LLaMA 3.1 8B consistently generated the `event_type` key, and a valid event type value `{flood, burns_scars}`, even when the query only referenced `crops` and in no way referenced `flood` or `burn_scars`. Its answers thereby passed key- and event-type-checks except `consistent_event_type`, which requires consistency between the user query and the generated `event_type` value. In contrast, answers generated with 405B most often excluded the `event_type` key entirely, contrary to the instructions, but included an error message per the instructions. Hence, 405B's overall score for `keys_required_present`, `valid_event_type`, `event_type_equivalent`, and `consistent_event_type` was 68.2%, matching 8B's score on `consistent_event_type`.

Further enlightening, were the LLM-as-a-judge generated critiques from *Date Consistency* evaluation traces. First, the critiques were not restricted to dates as instructed, but also examined other aspects of the answer, including event types and error messages. For example, a common error message in LLaMA 3.1 405B-generated answers was, "Event type not specified". A respective LLM-judge critique correctly deeming the answer inconsistent was, "...the response also contains an error message stating that the event type is not specified". Among the minority of critiques for 405B answers, unrelated to event or error messages, 2 flagged truly inconsistent dates, 1 reported "somewhat consistent" instead of a binary e.g. True/False (appropriately reflecting the query ambiguity), and 1 was due to a common type of parsing error when working with LLM-generated text (parsing `'consistent.'`). For LLaMA 3.1 8B-generated answers, all answers failing `consistent_date` were due to `flood` or `burn_scars` being used in place of `crops` as noted in the LLM-judge's critique. While not the originally intended role for the LLM-judge, the high accuracy of its critiques demonstrated potential for improving both generation and evaluation quality. Incorporating critique via iterative self-refinement (Madaan et al., 2024), and other approaches balancing generation versus verification complexity (Davis et al., 2024) for different tasks, is described in Section 3.3. However, coupling generated critiques with deterministic heuristics provided essential grounding.

In this scenario, `date_equivalent` (Figure 1, Left) better isolated temporal reference interpretation performance. Figure 1 (Right) shows the percentage of answers from LLaMA 3.1 405B failing this check in 3 categories. The predominant cause of failures (17%) was failing to follow instructions (or arguably, lack of instruction specificity), often with respect to date ranges. For example, multiple answers in this category listed every single day in the range rather than just the first and optionally last date of the range as instructed. A few of these, while ignoring at least 1 other instruction, also included dates in the future; a restriction (assumption) not stated in the instructions. Answers in the second largest category (3%) contained clear misinterpretations or miscalculations. For example, listing the dates for *this season last year* was counted as a clear misinterpretation of "last season". Failures where the golden answer or the generated answer appeared equally reasonable, with respect to natural language ambiguity in the query, constitute the final category (2%). For example, in (descriptive) North American English, the start of "this week" can refer to either Sunday or Monday without additional context. Similarly, an argument could be made for interpreting the start date of "this spring" as the first date of the first month of spring (March 1) according to the prompt instructions, or the date of the Vernal Equinox (March 20th) in the Northern Hemisphere... Example inputs and outputs are included in Appendix A.3.

This case study is like many others emphasizing the need for clear specification mechanisms in developing maintainable, debuggable, etc. Agentic and Compound AI systems (Stoica et al., 2024; Zaharia et al., 2024). Furthermore, it highlights the importance of performance regression tracking and continuous evaluation systems that account for nuances from LLM behavior. In this vein, Section 3.2 presents the results of employing a leading open-source framework, DSPy, for programmatic prompt optimization and systematic evaluation. See Section 2.6 regarding implementation.

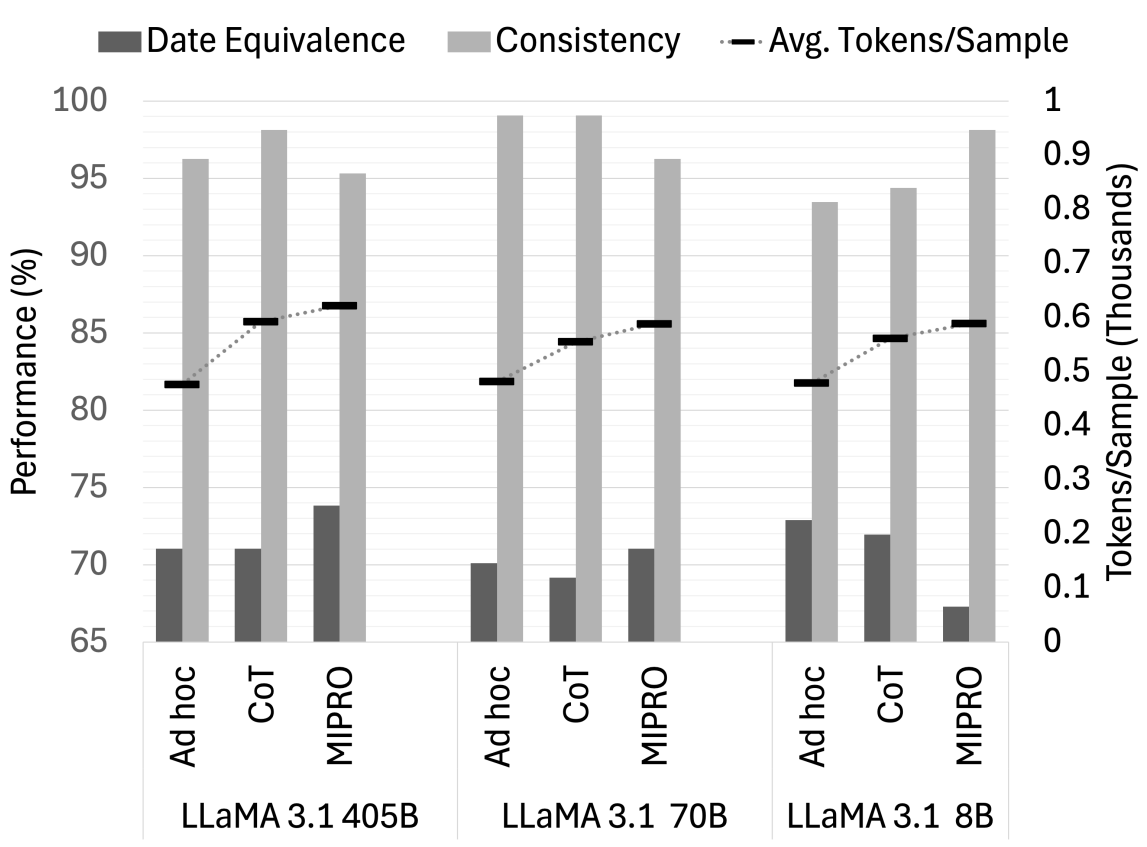

Figure 2: Date equivalence and LLM-judged consistency performance (left axis, 65-100%) across prompting techniques and LLaMA 3.1 variants, with tokens-spend as total input+output tokens per sample – query-answer pair (right axis, thousands).

## 3.2 Programmatic Prompt Optimization

We systematically explored the performance impact, cost (token usage), and throughput (queries/s) across prompting techniques and LLaMA 3.1 variants, 405B, 70B, and 8B. Specified using DSPy, we evaluated a simplified zero-shot version of the previously handwritten prompt ("Ad Hoc") alongside labeled few-shot (Brown et al., 2020), Chain of Thought (CoT) (Wei et al., 2022), and MIPRO (Opsahl-Ong et al., 2024). Overall, the metrics most significantly influenced were *Date Equivalence* and LLM-judged consistency, using *Ad Hoc*, CoT, and MIPRO, as shown in Figure 2.

*Ad Hoc* is a more concise version of the original prompt as a DSPy `Signature`. DSPy automatically injects standardized prompt-formatting, which makes the expected output fields (fields for auto-completion) and expected output format abundantly clear. Its performance was nearly on par with the more optimized prompts.

Without adding development complexity, Chain of Thought (CoT) offered reasonably competitive performance along with explanations (rationale) useful for development-time analysis and deployment-time transparency. We also evaluated a few-shot approach, in which examples were randomly sampled from the curated data set and excluded from evaluation. However, across repeated samplings and varying sample pool sizes, the per-formance did not significantly differ from that achieved with CoT alone.

Lastly, we attempted to optimize the zero-shot CoT instructions with MIPRO (Opsahl-Ong et al., 2024) as implemented in DSPy 2.5. We varied the hyperparameters, specifically the number of candidates, trials, mini-batch, and candidate-validation samples, using DSPy's automatic "light" and "heavy" configurations. While "heavy" optimization cost $1.8\times$ as many tokens as "light" optimization, the resulting prompts and their performance were not significantly different. The best-performing prompt from MIPRO optimization and the CoT prompt differed little. MIPRO optimization introduced a phrase at the beginning, "optimized for geographical events and phenomena", and at the end, "Consider the specific topics covered in the dataset, such as crop types, flooding, and burn scars, when generating the json mapping". The only other difference was the removal of newline characters unrelated to the desired output structure. The most significant performance impact was for LLaMA 3.1 8B, where consistency improved by $4.67\%$. However, date equivalence performance also dropped $5.61\%$, and the optimization process used approximately $1.5$ million tokens over $2$ thousand generations ($750$ tokens/trial). Since the resulting prompt can easily be saved and loaded, this cost can be amortized with use. Then again, the optimization needs to be repeated whenever the model choice or the pipeline changes significantly.

Figure 2 (right axis) also shows the average token-spend (post-optimization) per query across techniques and model variants as prompt tokens (input) plus tokens generated (output). Average token-spend gradually increased with the sophistication of the prompt and not necessarily with model scaling. Examining runtime speed improvements across model sizes, we observed on average $1.1\times$ and $1.8\times$ improvements in throughput over LLaMA 3.1 405B from LLaMA 3.1 70B and 8B, respectively. In conclusion, CoT offered a sweet spot with respect to development- and deployment-time token-spend, accuracy, and increased transparency, and LLaMA 3.1 8B offered sufficiently competitive performance and latency-cost savings as to be the focal point of ongoing optimization efforts, including those in Section 3.3.

### 3.3 Inference-Time Scaling Optimizations

Given the accuracy-speed potential of LLaMA 3.1 8B with CoT shown in Section 3.2, this section examines the potential of inference-time scaling for addressing the remaining *Date Equivalence* and *Date Consistency* gaps (6% and 28%, respectively). To establish baseline cost-accuracy gains for this direction, we begin with 1-step iterative self-refinement (Madaan et al., 2024) and a simple task decomposition with mixed models, illustrated in Figure 3. In contrast to the previously evaluated single-step CoT (Figure 3(i)), self-refinement adds an additional inference layer (Figure 3(ii)) in which $\text{Model}_b$ is prompted to consider the answer generated by $\text{Model}_a$ according to the same criteria and refine the response if necessary. In general, $\text{Model}_a$ and $\text{Model}_b$ (corresponding to generator and refiner) may be the same or different models. The task decomposition approach (Figure 3(iii)) is a mixture-of-experts inspired approach. The most challenging task, date interpretation, is separated from the the tasks with high success rates by prompting models $a$ and $b$ to generate answers only for their respective subtasks. The final layer prompts $\text{Model}_c$ to synthesize the generated answers into a single coherent and refined answer. In general, model selection for $a$, b, and $c$ may differ as parameterized in our code.

Figure 4 highlights the results employing LLaMA 3.1 8B for models $a$ and $c$, and varying model $b$ between LLaMA 3.1 8B and 405B. All methods improve LLM-judged consistency by 1.9 to 3.7 points. However, only Split-Generate-Synthesize with LLaMA 3.1 405B for date extraction ($\text{Model}_b$ in Figure 3.$iii$, *SGS 8B+405B* in Figure 4), improves both consistency and *Date Equivalence*, by 0.94 and 1.87 points respectively, over single-step CoT. The cost in average tokens spent processing each query on the other hand was $2.6\times$ that of CoT. Single-step iterative self-refinement and SGS with LLaMA 3.1 8B for Models $a$, $b$, and $c$ attained slightly higher consistency (within 1 to 2 points) at the cost of lower *Date Equivalence* (4.6 to 6.6 points lower). Scaling the number of iterations of self-refinement or the task decomposition could improve performance further. However, the cost would be multiplicative, and for a single iteration it is already $2\times$ to $2.6\times$ the cost of CoT. The development (prompt tuning, model selection, etc.) and maintenance cost of each inference layer also presents a drawback.

## 4 Discussion and Future Work

Highlighted in the context of our implementation and experimentation (Section 2.6-3) was the need for multi-dimensional quality optimization; measuring accuracy, cost, latency simultaneously across techniques; and supporting a spectrum of techniques from prompt optimization to inference-time scaling, in both the programming (specification) and evaluation system. Recent frameworks may address some of the gaps noted in Section 2.6, such as Archon for inference-time scaling (Saad-Falcon et al., 2024). However, managing the complexity and unifying these and other aspects of Agentic and Compound AI system development is still open as new methods for optimizing Compound AI systems emerge daily.

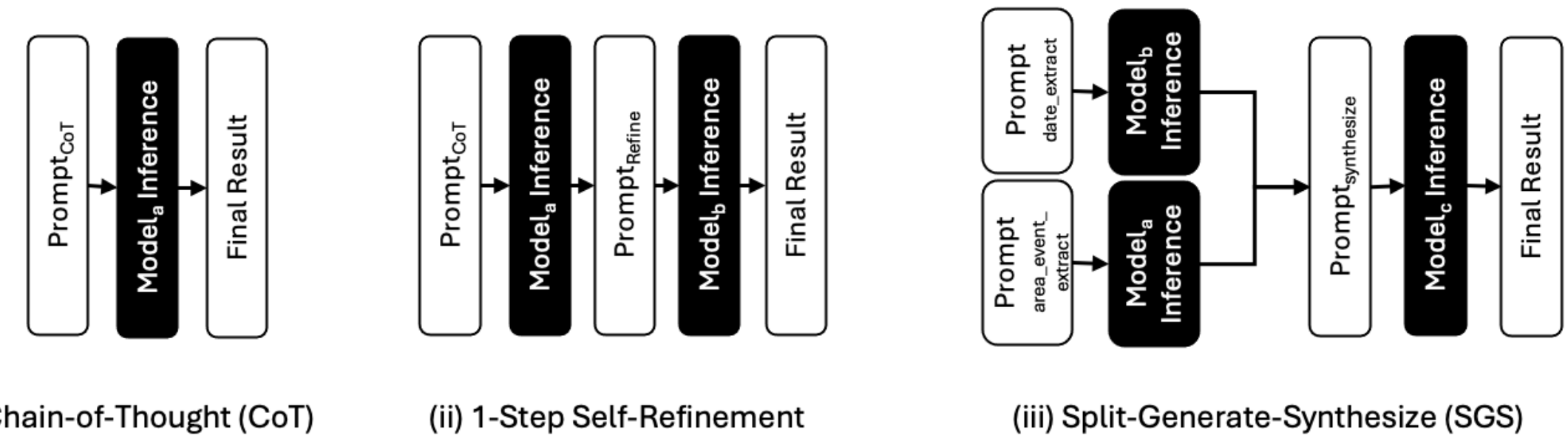

Figure 3: Illustration of the inference scaling technique baselines (ii-iii) alongside single-step Chain of Thought (i) used to establish baseline costs and gains from inference-time scaling. White boxes indicate modifiable steps, inputs or outputs. Black boxes are treated as opaque, immutable processes.

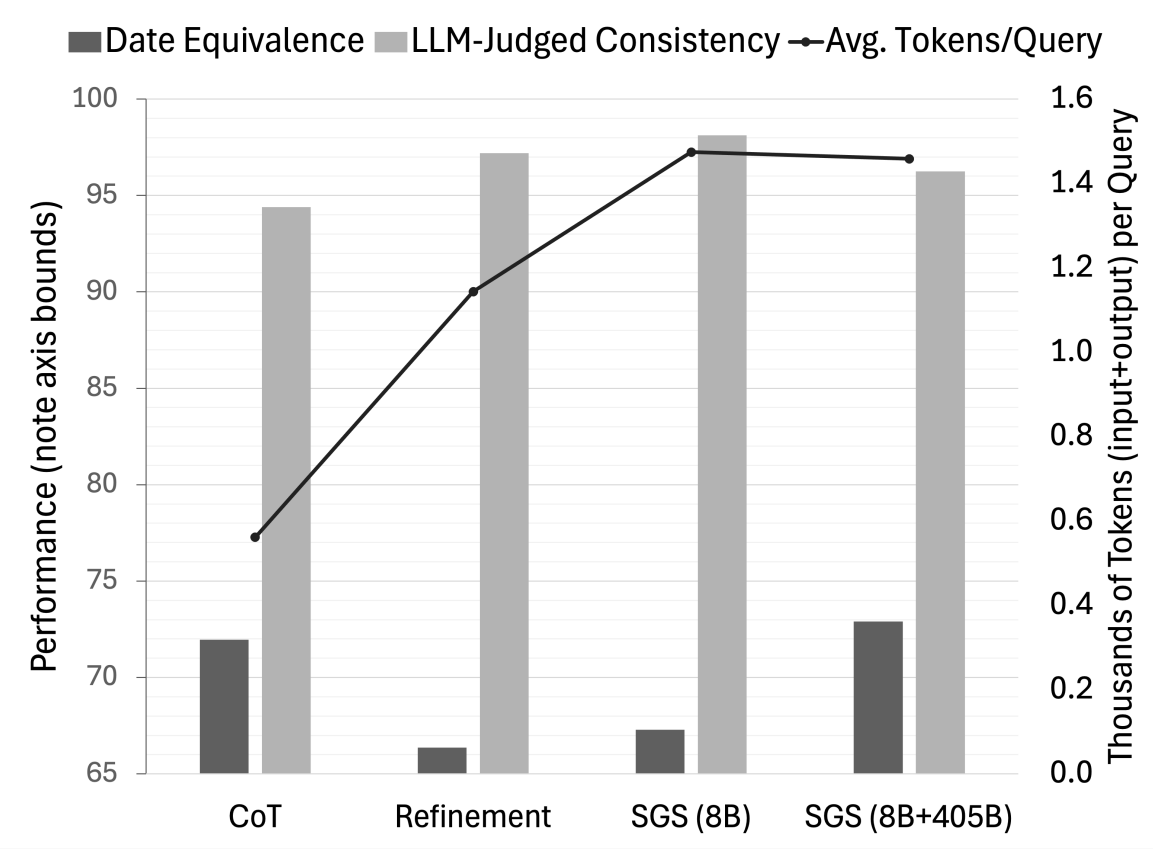

Figure 4: Performance on Date Equivalence and LLM-Judged Consistency (left axis, 65-100%) and tokens per query (right axis, thousands) across optimization techniques described in Section 3.3. LLaMA 3.1 8B was used exclusively, except for *SGS (8B+405B)*, which also employs 405B.

Moreover, we would like to understand optimization trade-offs not only at manually-triggered points in time, but also over time as the evaluation data changes and new optimization techniques become available. The effectiveness of LLM-assisted evaluation data vetting described in Section 2.3 suggests a scalable approach for curating more data over time. Methods ensuring the curated data is error-free and representative of user interactions, however requires further research. For efficiency and scalability, we envision a system that leverages LLM-infused batch-processing methods (e.g. LOTUS (Patel et al., 2024)) for continuous evaluation and that automate failure diagnoses (Section 3.1) with new and traditional ML techniques (e.g. classification). In addition to generalizing our programming and evaluation system, our ongoing work includes evaluating human-in-the-loop and co-piloting approaches for improving accuracy. The above and other directions are expanded in Appendix Section A.2.

## 5 Summary and Conclusions

This study was a first exploring the potential of Large Language Models (LLMs) to democratize earth observation data analysis (Section 1) with open models and open-source software. We reformulated the problem as a Named Entity Recognition task and built an optimization and evaluation framework (Section 2). We established baseline accuracy, cost, and latency trade-offs across model scaling, prompt optimization, and certain inference-time scaling techniques (Section 3). The combination of relatively small general purpose models (e.g. LLaMA 3.1 8B), simple techniques (e.g. Chain of Thought), and principled software engineering stood out. Interpreting relative spatiotemporal references from natural language queries with high accuracy, consistency, and safeguards remains a priority for ongoing work. Furthermore, we related our findings to the broader field, and enumerated requirements and opportunities for Agentic and Compound AI programming, evaluation, and opti-mization systems more generally (Sections 4, A.2). The assets developed for this study provide a foundation for tackling the research and deployment challenges of this important application. We welcome contributors. Code is available at https://github.com/NASA-IMPACT/EO-via-NLP. The data is available at https://huggingface.co/datasets/nasa-impact/EO-via-NLP.

## 6 Limitations

### 6.1 Safety

As noted in the introduction, the scope of this study is limited to the accuracy, cost, latency, and maintainability of a LLM-powered solution. Safety, however, is also critical to a deployable solution. This includes safe-guards to prevent hallucination, hijacking, DoS attacks, and other malicious attacks and behaviors that could negatively impact users, providers, or the system. It also includes addressing potential biases embedded within LLMs. While outside the scope of this initial study, our ongoing work prioritizes guardrailing mechanisms and rigorous grounding of the system's responses to ensure factual accuracy and mitigate environmental, political, and societal biases.

### 6.2 Model and Technique Selection & Search

Agentic and Compound AI system development presents a large optimization space of co-optimizing model selection, prompts, model parameters such as temperature, et cetera. For tractability, we started with the LLaMA 3.1 family, as well-vetted open and general purposes models, varied model sizes, explored prompt optimization strategies, and set a baseline for exploring inference-time scaling techniques. The inference-time scaling baseline designs increased width by 1 (task decomposition from 1 to 2) and depth by 1 (single iteration self-refinement, and separately synthesis) as illustrated in Figure 3. Increasing the depth and width further may very well achieve higher accuracy. The relatively high cost (token-spend) of doing so ($\geq 2\times$ in our study) may be addressed with more concise models or alternative means of limiting output. One drawback of limiting output we encountered was that it cut rationale, valuable for system transparency and debuggability; however alternate methods for limiting output may simultaneously address this drawback. Evaluating ensembles of a variety of small models for balancing accuracy and cost (Chen et al., 2023) is another welcome extension of this study. Lastly, the new generation of "reasoning" models, such as Deepseek-R1 (Wang et al., 2025) and subsequent variants, may improve performance on spatiotemporal reference interpretation. In general, any change of models requires re-examination and re-optimizing of system instructions and hyperparameters, among many other aspects of the implementation. This is both a limitation of the study presented and an opportunity for future work as noted in Sections 4 and A.2.

### 6.3 Evaluation Data Set Limitations

The ambiguity inherent in natural language queries allows for multiple valid answers. However, the current data set presents golden queries and answers in a 1:1 ratio. There is an opportunity to expand the answer set, essentially loosening the optimization constraints in a helpful way.

Query and answer-generation complexity varies significantly and non-uniformly across the data set. Our discussion of the subtask complexity and variation (Section 2.4) may be applied to generating additional evaluation data of a target complexity, and also to classifying the difficulty of a given QA pair. This remains future work.

Lastly, multiple dates or date ranges from an individual query are supported by the evaluation procedures. However, multiple areas or event types in a single query are not represented in the evaluation data set or procedure. This might be natural to express in human language, but would require significant changes in the surrounding system. For example, tasking the router with creating multiple queries and rephrasing, or increasing specificity with human-in-the-loop are reasonable approaches and directions for future work. However, this study analyzed the potential of supporting single-topic queries with LLMs as a first step.

## Acknowledgments

This work is supported by NASA Grant 80MSFC22M004. We thank Jared Quincy Davis and Paul Castro for their early support.

## References

Common Metadata Repository. https://www.earthdata.nasa.gov/about/esdis/eosdis/cmr. [Online; last accessed 11-Feb-2025].

Tom Brown, Benjamin Mann, Nick Ryder, Melanie Subbiah, Jared D Kaplan, Prafulla Dhariwal, Arvind

Neelakantan, Pranav Shyam, Girish Sastry, Amanda Askell, et al. 2020. Language models are few-shot learners. *Advances in Neural Information Processing Systems*, 33:1877–1901.

Harrison Chase. 2022. LangChain.

Lingjiao Chen, Matei Zaharia, and James Zou. 2023. Frugalgpt: How to use large language models while reducing cost and improving performance. *Preprint*, arXiv:2305.05176.

Wei-Lin Chiang, Lianmin Zheng, Ying Sheng, Anastasios Nikolas Angelopoulos, Tianle Li, Dacheng Li, Banghua Zhu, Hao Zhang, Michael Jordan, Joseph E Gonzalez, et al. 2024. Chatbot arena: An open platform for evaluating llms by human preference. In *Forty-first International Conference on Machine Learning*.

Jared Quincy Davis, Boris Hanin, Lingjiao Chen, Peter Bailis, Ion Stoica, and Matei Zaharia. 2024. Networks of networks: Complexity class principles applied to compound ai systems design. *CoRR*.

Lisa Dunlap, Krishna Mandal, Trevor Darrell, Jacob Steinhardt, and Joseph E Gonzalez. 2024. Vibecheck: Discover and quantify qualitative differences in large language models. *arXiv preprint arXiv:2410.12851*.

IBM and NASA. 2023. Ibm-nasa prithvi models family.

Johannes Jakubik, Sujit Roy, CE Phillips, Paolo Fraccaro, Denys Godwin, Bianca Zadrozny, Daniela Szwarcman, Carlos Gomes, Gabby Nyirjesy, Blair Edwards, et al. 2023. Foundation models for generalist geospatial artificial intelligence. *CoRR*.

Omar Khattab, Arnav Singhvi, Paridhi Maheshwari, Zhiyuan Zhang, Keshav Santhanam, Sri Vardhamanan, Saiful Haq, Ashutosh Sharma, Thomas T. Joshi, Hanna Moazam, Heather Miller, Matei Zaharia, and Christopher Potts. 2024. Dspy: Compiling declarative language model calls into self-improving pipelines.

Aman Madaan, Niket Tandon, Prakhar Gupta, Skyler Hallinan, Luyu Gao, Sarah Wiegreffe, Uri Alon, Nouha Dziri, Shrimai Prabhumoye, Yiming Yang, et al. 2024. Self-refine: Iterative refinement with self-feedback. *Advances in Neural Information Processing Systems*, 36.

Meta AI. 2024. The llama 3 herd of models.

Microsoft Open Source, Matt McFarland, Rob Emanuele, Dan Morris, and Tom Augspurger. 2022. microsoft/planetarycomputer: October 2022.

D. J. Newman and ESDIS Standards Coordination Office ESCO. 2023. SpatioTemporal Asset Catalogs (STAC). NASA Earth Science Data and Information System Standards Coordination Office.

Krista Opsahl-Ong, Michael J Ryan, Josh Purtell, David Broman, Christopher Potts, Matei Zaharia, and Omar Khattab. 2024. Optimizing instructions and demonstrations for multi-stage language model programs. *arXiv preprint arXiv:2406.11695*.

Liana Patel, Siddharth Jha, Carlos Guestrin, and Matei Zaharia. 2024. Lotus: Enabling semantic queries with llms over tables of unstructured and structured data. *arXiv preprint arXiv:2407.11418*.

Jon Saad-Falcon, Adrian Gamarra Lafuente, Shlok Natarajan, Nahum Maru, Hristo Todorov, Etash Guha, E. Kelly Buchanan, Mayee Chen, Neel Guha, Christopher Ré, and Azalia Mirhoseini. 2024. Archon: An architecture search framework for inference-time techniques.

Arnav Singhvi, Manish Shetty, Shangyin Tan, Christopher Potts, Koushik Sen, Matei Zaharia, and Omar Khattab. 2024. Dspy assertions: Computational constraints for self-refining language model pipelines. *Preprint*, arXiv:2312.13382.

Ion Stoica, Matei Zaharia, Joseph Gonzalez, Ken Goldberg, Hao Zhang, Anastasios Angelopoulos, Shishir G Patil, Lingjiao Chen, Wei-Lin Chiang, and Jared Q Davis. 2024. Specifications: The missing link to making the development of llm systems an engineering discipline. *arXiv preprint arXiv:2412.05299*.

Yixuan Wang, Xiaohui Li, Yiming Zhang, Jiawei Zhang, Jun Zhu, Jian Chen, and Wei Li. 2025. Deepseek-r1: Incentivizing reasoning capability in llms via reinforcement learning. *arXiv preprint arXiv:2501.12948*.

Jason Wei, Xuezhi Wang, Dale Schuurmans, Maarten Bosma, Fei Xia, Ed Chi, Quoc V Le, Denny Zhou, et al. 2022. Chain-of-thought prompting elicits reasoning in large language models. *Advances in Neural Information Processing Systems*, 35:24824–24837.

Matei Zaharia, Omar Khattab, Lingjiao Chen, Jared Quincy Davis, Heather Miller, Chris Potts, James Zou, Michael Carbin, Jonathan Frankle, Naveen Rao, and Ali Ghodsi. 2024. The shift from models to compound ai systems. `https://bair.berkeley.edu/blog/2024/02/18/compound-ai-systems/`.

Lianmin Zheng, Wei-Lin Chiang, Ying Sheng, Siyuan Zhuang, Zhanghao Wu, Yonghao Zhuang, Zi Lin, Zhuohan Li, Dacheng Li, Eric Xing, Hao Zhang, Joseph E Gonzalez, and Ion Stoica. 2023. Judging llm-as-a-judge with mt-bench and chatbot arena. In *Advances in Neural Information Processing Systems*, volume 36, pages 46595–46623. Curran Associates, Inc.

# A Appendix

## A.1 Additional Experimental Details

1. Results presented were collected using IBM WatsonX model serving via DSPy `>=2.5`

LiteLLM integration [3] [4].

2. String equivalence measurements were computed using RapidFuzz 3.10 [5], with a normalized insertion-deletion similarity cutoff of 0.7.

### A.2 Extended Future Work Discussion

**Semi-Automatic Evaluation-Data Generation**. Our experience developing an evaluation data set led to the following viewpoint: maintaining a balanced (representative) evaluation data set for system evaluation over time, has the potential and ought to be a continuously-running and semi-automatic process including (a) ingesting data from amenable users and data contributors, (b) filtering and curating this data semi-automatically with machine learning methods, including but not limited to employing LLMs as noted in Section 2.3, and (c) manually approving the final, minimal set of proposed changes per version update, mirroring software release cycles. Mature storage and telemetry systems make part of this problem purely an engineering effort. Automatic curation, ensuring evaluation data is error-free and representative of user interactions, however is an interesting problem requiring further research.

**Programming Frameworks**. Agentic and Compound AI system developers have a wide-range of emerging programming frameworks to choose from, from purpose-built frameworks (e.g. DSPy for prompt-optimization (Khattab et al., 2024)) to catch-all frameworks (e.g. LangChain (Chase, 2022)). Our initial prototype used LangChain; we shared our experience subsequently moving from handwritten- to programmatic-prompt optimization with DSPy. Newer frameworks may address some of the gaps we noted in Section 2.6, e.g. Archon (Saad-Falcon et al., 2024). However, we believe there is still more work to do in unifying the aspects of Compound AI system development we discussed (model selection, prompt optimization, inference-time scaling, model hyperparamter tuning) as well as others e.g. safety, or providing true interoperability across targeted solutions. This is necessarily open research as new ways of optimizing Compound AI systems are emerging daily.

[3] https://dspy.ai/learn/programming/language_models/?h=litellm#__tabbed_1_6. Last accessed Feb. 7, 2025.

[4] https://docs.litellm.ai/docs/providers/watsonx. Last accessed Feb. 7, 2025.

[5] https://rapidfuzz.github.io/RapidFuzz/Usage/fuzz.html#ratio. Last accessed Feb. 7, 2025.

**Evaluation Frameworks**. Section 2.6 described the advantages and limitations of DSPy and its evaluation subsystem for our use-case; we implemented a framework sufficient for our use-case in Python with DSPy, and generalizing it is part of ongoing work. Highlighted in particular was the need for general multi-objective optimization; measuring accuracy, cost, latency simultaneously across techniques; and supporting a spectrum of techniques from prompt optimization to inference-time scaling, in both the programming (specification) and evaluation system. More broadly, we would like to understand these trade-offs at not only manually-triggered points in time, but over time as the evaluation data changes and new optimization techniques become available. Such continuous evaluation could mirror automated CI/CD pipelines, common for other software. Additionally, there exists an opportunity to accelerate continuous evaluation via emerging methods for LLM-integrated offline bulk-processing with new frameworks (Patel et al., 2024), as well as employing traditional ML techniques (e.g. classification) for diagnosing failures as in Section 1.

**Safety**. As noted in the introduction, the scope of this study is limited to the accuracy, cost, latency, and maintainability of a LLM-powered solution. Safety, however, is also critical to a deployable solution. This includes safe-guards to prevent hallucination, hijacking, DoS attacks, and other malicious attacks that could negatively impact users, providers, or the system. It also includes addressing potential biases embedded within LLMs. Our ongoing work prioritizes guardrailing mechanisms and rigorous grounding of the system's responses to ensure factual accuracy and mitigate environmental, political, and societal biases.

**Human-in-the-Loop (HIL) and Copiloting**. This study focused on recognizing the minimally necessary set of parameters necessary for driving earth observation data analysis from a single natural language query. However, it may not be possible in general to determine exact intent from the initial query, due to for example, ambiguity inherent in natural language or missing information. Therefore, HIL may be most effective for achieving high accuracy (especially for temporal references) while avoiding resource waste (time, compute, etc.) from mis-speculation. A copiloting approach may provide greater workflow customizability. We are exploring these paths in the context of the overall system deployment.

**Model and Technique Search and Selection**. Section 6 noted limitations regarding model selection and the techniques examined. In particular, the new generation of "reasoning" models, such as Deepseek-R1 (Wang et al., 2025), may address the performance gap we observed for spatiotemporal reference interpretation. However, with any change of model, other aspects such as re-optimizing system instructions, temperature, inference-time scaling architecture, etc. need to simultaneously be re-examined. Solutions accelerating parts of this process are emerging; Archon (Saad-Falcon et al., 2024) for example translates hyperparameter tuning to this context. For initial model selection, aside from manual sampling and reading model cards, popular mechanisms for model selection are based on general academic benchmarks and crowdsourced rankings (Chiang et al., 2024; Dunlap et al., 2024). There is yet an opportunity to accelerate model selection process from the start according to task complexity, and benchmarks meaningful for specific application workflows.

### A.3 Date-Equivalence Failure Examples Using Llama 3.1 405B

Below are example traces supplementing the description in Section 3.1. In multiple examples, the generated answer includes every day in a range; for large ranges, dates in the middle are displayed with ellipses for brevity.

```
(60) Query: Provide the latest imagery of
     flooding in Houston, Texas, from this past
     Tuesday. Today is June 4, 2024.
(60) Generated Answer: {
    "area": "Houston, Texas",
    "date": "2024-06-02",
    "event_type": "flood",
    "error": ""
}
(60) Golden Answer: {"area": "Houston, Texas", "
     date": "2024-05-28","event_type":"flood"}

(63) Query: Highlight recent flooding events in
     the UK from this past Spring. Today is June
     4, 2024.
(63) Generated Answer: {
    "area": "UK",
    "date": ["2024-03-01", "2024-03-02", \ldots
        "2024-05-30", "2024-05-31"],
    "event_type": "flood"
}
(63) Golden Answer: {"area": "UK", "date":
     "2024-03-01","event_type":"flood"}
----
(66) Query: Display the latest flooding events
     in Toronto, Canada, from last month. Today
     is June 4, 2024.
(66) Generated Answer: {
    "area": "Toronto, Canada",
    "date": ["2024-05-01", "2024-05-02", \ldots
        "2024-05-30", "2024-05-31"],
    "event_type": "flood",
    "error": ""
}
(66) Golden Answer: {"area": "Toronto, Canada",
     "date": "2024-05-01","event_type":"flood"}
----
(67) Query: Show satellite imagery of burn scars
     in Morocco from this year. Today is June 4,
     2024.
(67) Generated Answer: {
    "area": "Morocco",
    "date": ["2024-01-01", "2024-01-02", \ldots
        "2024-06-03", "2024-06-04"],
    "event_type": "burn_scars"
}
(67) Golden Answer: {"area": "Morocco", "date":
     "2024-01-01","event_type":"burn_scars"}
----
(68) Query: Can you find crop types in Kansas as
     of the last 30 days? Today is June 4, 2024.
(68) Generated Answer: {
    "area": "Kansas",
    "date": ["2024-05-05", "2024-05-06", \ldots,
        "2024-06-03", "2024-06-04"],
    "error": "Event type not specified"
}
(68) Golden Answer: {"area": "Kansas", "date":
     "2024-05-04","event_type":"crops"}
----
(69) Query: Provide images of recent flooding in
     Cairo, Egypt, from the past week. Today is
     June 4, 2024.
(69) Generated Answer: {
    "area": "Cairo, Egypt",
    "date": ["2024-05-28", "2024-05-29", \ldots
        "2024-06-03"],
    "event_type": "flood",
    "error": ""
}
(69) Golden Answer: {"area": "Cairo, Egypt", "
     date": "2024-05-28","event_type":"flood"}
----
(73) Query: Find burn scars in the Kalahari
     Desert from the past three months. Today is
     June 4, 2024.
(73) Generated Answer: {
    "area": "Kalahari Desert",
    "date": ["2024-03-01", "2024-04-01",
        "2024-05-01", "2024-06-01"],
    "event_type": "burn_scars",
    "error": ""
}
(73) Golden Answer: {"area": "Kalahari Desert",
     "date": "2024-03-04","event_type":"
     burn_scars"}
----
(74) Query: Can you show crop types in the
     Netherlands observed last weekend? Today is
     June 4, 2024.
(74) Generated Answer: {
    "area": "Netherlands",
    "date": ["2024-06-01", "2024-06-02",
        "2024-06-03"],
    "error": "Event type not specified"
}
(74) Golden Answer: {"area": "Netherlands", "
     date": "2024-06-01","event_type":"crops"}
```

----
(75) Query: Highlight flooding events in Lima, Peru, from this Spring. Today is June 4, 2024.
(75) Generated Answer: {
"area": "Lima, Peru",
"date": ["2024-03-20", "2024-03-21", \ldots "2024-06-04"],
"event_type": "flood",
"error": ""
}
(75) Golden Answer: {"area": "Lima, Peru", "date": "2024-03-01","event_type":"flood"}
----
(79) Query: Find burn scars in the Andes Mountains from last season. Today is June 4, 2024.
(79) Generated Answer: {
"area": "Andes Mountains",
"date": ["2023-12-01", "2023-12-31", "2024-01-01", "2024-01-31", "2024-02-01", "2024-02-28", "2024-03-01", "2024-03-31"],
"event_type": "burn_scars",
"error": ""
}
(79) Golden Answer: {"area": "Andes Mountains", "date": "2024-03-01","event_type":"burn_scars"}
----
(80) Query: Display the latest crop types in Israel observed this Friday. Today is June 4, 2024.
(80) Generated Answer: {
"area": "Israel",
"date": "2024-06-07",
"error": "Event type not specified"
}
(80) Golden Answer: {"area": "Israel", "date": "2024-05-31","event_type":"crops"}
----
(84) Query: Display recent flooding in New York City from the last 48 hours. Today is June 4, 2024.
(84) Generated Answer: {
"area": "New York City",
"date": ["2024-06-02", "2024-06-03", "2024-06-04"],
"event_type": "flood",
"error": ""
}
(84) Golden Answer: {"area": "New York City", "date": "2024-06-02","event_type":"flood"}
----
(85) Query: Find the most recent burn scars in Fiji from the past year. Today is June 4, 2024.
(85) Generated Answer: {
"area": "Fiji",
"date": ["2023-01-01", "2023-12-31"],
"event_type": "burn_scars",
"error": ""
}
(85) Golden Answer: {"area": "Fiji", "date": "2023-06-04","event_type":"burn_scars"}
----
(87) Query: Highlight flooding events in Paris, France, from last weekend. Today is June 4, 2024.
(87) Generated Answer: {
"area": "Paris, France",
"date": ["2024-06-01", "2024-06-02", "2024-06-03"],
"event_type": "flood",
"error": ""
}
(87) Golden Answer: {"area": "Paris, France", "date": "2024-06-01","event_type":"flood"}
----
(88) Query: Display satellite imagery of burn scars in Tasmania from this summer. Today is June 4, 2024.
(88) Generated Answer: {
"area": "Tasmania",
"date": ["2023-12-01", "2023-12-02", \ldots "2024-03-31"],
"event_type": "burn_scars",
"error": ""
}
(88) Golden Answer: {"area": "Tasmania", "date": "2024-06-01","event_type":"burn_scars"}
----
(89) Query: Can you find crop types in Egypt observed last Thursday? Today is June 4, 2024.
(89) Generated Answer: {"area": "Egypt", "error": "Event type not found. Also, 'crop types' is not a valid event type. It should be either 'burn_scars' or 'flood'."}
(89) Golden Answer: {"area": "Egypt", "date": "2024-05-30","event_type":"crops"}
----
(91) Query: Display burn scars in Nova Scotia from last month. Today is June 4, 2024.
(91) Generated Answer: {
"area": "Nova Scotia",
"date": ["2024-05-01", "2024-05-02", \ldots, "2024-05-31"],
"event_type": "burn_scars",
"error": ""
}
(91) Golden Answer: {"area": "Nova Scotia", "date": "2024-05-01","event_type":"burn_scars"}
----
(92) Query: Highlight crop types in Argentina seen this past weekend. Today is June 4, 2024.
(92) Generated Answer: {
"area": "Argentina",
"date": "2024-06-02",
"error": "Event type not specified"
}
(92) Golden Answer: {"area": "Argentina", "date": "2024-06-01","event_type":"crops"}
----
(94) Query: Show the most recent burn scars in the Pyrenees from this week. Today is June 4, 2024.
(94) Generated Answer: {
"area": "Pyrenees",
"date": ["2024-06-01", "2024-06-02", "2024-06-03", "2024-06-04"],
"event_type": "burn_scars",
"error": ""
}
(94) Golden Answer: {"area": "Pyrenees", "date": "2024-06-02","event_type":"burn_scars"}
----
(96) Query: Provide images of flooding in Lagos,

```
     Nigeria, from the last 72 hours. Today is
    June 6, 2024.
(96) Generated Answer: {
   "area": "Lagos, Nigeria",
   "date": ["2024-06-04", "2024-06-05",
       "2024-06-06"],
   "event_type": "flood",
   "error": ""
}
(96) Golden Answer: {"area": "Lagos, Nigeria", "
    date": "2024-06-04","event_type":"flood"}
----
(97) Query: Show burn scars in the Everglades
    from last season. Today is June 4, 2024.
(97) Generated Answer: {
   "area": "Everglades",
   "date": ["2023-06-01", "2023-08-31",
       "2023-09-01", "2023-11-30", "2023-12-01",
        "2023-12-31"],
   "event_type": "burn_scars",
   "error": ""
}
(97) Golden Answer: {"area": "Everglades", "date
    ": "2024-03-01","event_type":"burn_scars"}
----
(102) Query: Provide satellite images of
    flooding in Kyoto, Japan, from this year.
    Today is June 4, 2024.
(102) Generated Answer: {
   "area": "Kyoto, Japan",
   "date": ["2024-01-01", "2024-02-01",
       "2024-03-01", "2024-04-01", "2024-05-01",
        "2024-06-01"],
   "event_type": "flood",
   "error": ""
}
(102) Golden Answer: {"area": "Kyoto, Japan", "
    date": "2024-01-01","event_type":"flood"}
----
(105) Query: Highlight recent flooding events in
     Bangkok, Thailand, from the past week.
    Today is June 4, 2024.
(105) Generated Answer: {
   "area": "Bangkok, Thailand",
   "date": ["2024-05-28", "2024-05-29", \ldots,
       "2024-06-02", "2024-06-03"],
   "event_type": "flood",
   "error": ""
}
(105) Golden Answer: {"area": "Bangkok, Thailand
    ", "date": "2024-05-28","event_type":"flood
    "}
----
(106) Query: Show burn scars in the Amazon from
    this Spring. Today is June 4, 2024.
(106) Generated Answer: {
   "area": "Amazon",
   "date": ["2024-03-20"],
   "event_type": "burn_scars",
   "error": ""
}
(106) Golden Answer: {"area": "Amazon", "date":
    "2024-03-01","event_type":"burn_scars"}
----
```